\documentclass[12pt]{iopart}
\usepackage{graphicx}
\usepackage{color}
\usepackage{hyperref}

\begin{document}

\title{Refined analysis on the parton distribution functions of the proton}

\author{X G Wang, A W Thomas}
\address{CSSM and ARC Centre of Excellence for Particle Physics at the Terascale, Department of Physics, University of Adelaide SA 5005 Australia}
\eads{\mailto{xuan-gong.wang@adelaide.edu.au}, \mailto{anthony.thomas@adelaide.edu.au}}

\begin{abstract}
We explore the application of a two-component model of proton structure functions in the analysis of  deep-inelastic scattering (DIS) data at low $Q^2$ and small $x$. 
This model incorporates both vector meson dominance and the correct photo-production limit.
The CJ15 parameterization is applied to the QCD component, in order to take into account effects of order $1/Q^2$ effects, such as target mass corrections and higher twist contributions.
The parameters of the leading twist parton distribution functions and higher twist coefficient functions are determined by fitting deep inelastic scattering data.
The second moments of the parton distribution functions are extracted and compared with other global fits and lattice determinations.
\end{abstract}

\noindent{\it Keywords\/}: Parton distribution functions, vector meson dominance, deep inelastic scattering

\submitto{\jpg}

\maketitle


\section{Introduction}
Understanding the inner structure of the nucleon remains one of the most challenging tasks in modern particle and nuclear physics. 
It is of special interest how the nucleon's momentum and spin are divided among quarks and gluons.
Within quantum chromodynamics (QCD), this information can be accessed through parton distribution functions (PDFs). 
Specifically, the second moments of unpolarised PDFs are interpreted as the momentum fractions carried by partons. 

There are mainly two approaches to studying PDFs of nucleons. 
First, the PDFs are determined from global fits to the world data on deep inelastic scattering (DIS) and related hard scattering processes~\cite{MRST:1998, GRV:1998}.
One usually starts from a parametrisation of the initial PDFs at somehow low scale, and then evolves to high $Q^2$ region using the DGLAP equations.
Nowadays, with improvement in the precision and kinematic range of the experimental measurements from Jefferson Lab, HERA, RHIC, the Tevatron and the LHC, 
global fits have covered data over a broad range of Bjorken $x$ and four-momentum transfer $Q^2$~\cite{MMHT14, NNPDF:2015, CJ15}.

The second method is lattice QCD. Early lattice simulations were limited to the nucleon matrix elements of leading-twist (LT) local operators, which correspond to the low moments of PDFs.
Recent approaches to determining the $x$-dependent PDFs have been proposed, either in terms of quasi-PDFs~\cite{Ji:2013, Lin:2015}, or in terms of pseudo-PDFs~\cite{Radyushkin:2017, Orginos:2017}.

Efforts have been made to establish connections between these two approaches~\cite{Lin:2018}. 
The results from global fits are useful in validating current and future lattice simulations.  
On the other hand, the lattice results could be used to reduce the uncertainties in current global analysis of PDFs.
 
However, a surprising difference between global fits and lattice calculations of the moment of unpolarized flavor-singlet PDFs was observed recently. 
At $Q^2 = 4~{\rm GeV}^2$, the global fit determinations give
\begin{eqnarray}
\langle x \rangle^{{\rm exp}}_{u^+} &=& 0.352(12),\ \langle x \rangle^{{\rm exp}}_{d^+}=0.192(6),\nonumber\\
\langle x \rangle^{\rm exp}_{s^+} &=& 0.037(3), \ \langle x \rangle^{\rm exp}_g = 0.411(8),
\end{eqnarray}
while the latest lattice simulations from ETMC17 gave significantly larger results, albeit with large uncertainties~\cite{Alexandrou:2017},
\begin{eqnarray}
\langle x \rangle^{{\rm lat}}_{u^+} &=& 0.453(75),\ \langle x \rangle^{{\rm lat}}_{d^+}=0.259(74), \nonumber\\ 
\langle x \rangle^{{\rm lat}}_{s^+} &=& 0.092(41), \ \langle x \rangle^{\rm lat}_g = 0.267(35).
\end{eqnarray}
As discussed in Ref.~\cite{Lin:2018}, it may be that this discrepancy would be removed by taking into account the renormalisation properly and imposing momentum sum rule in lattice calculations.

However, the discrepancy may also originate from the global-fit side. The experimental results include both short and long distance physics. 
This can be seen from the flavor-singlet light-cone correlation functions which are obtained by Fourier transforming the empirical PDFs in momentum space.
The resulting distributions extend over large distances~\cite{Braun:1995},
reflecting primarily the partonic structure of the photon in the very small $x$ region, i.e., the virtual photon fluctuates into a $q\bar{q}$ pair~\cite{Vanttinen:1998}.
This effect is usually saturated by the vector meson dominance (VMD) model.

On the other hand, when $Q^2 \rightarrow 0$, $F_2$ must vanish linearly with $Q^2$ in order to give a finite photo-production cross section.
Such  behavior cannot be generated by DGLAP evolution to low-$Q^2$.

A two-component model to separate perturbative and non-perturbative contributions to the proton structure function was first proposed by Badelek and Kwiecinski~\cite{KB:1989,BK:1992}.
Incorporating the vanishing of $F_2$ in the $Q^2 = 0$ limit as well as the scaling behaviour at large $Q^2$,
the proton structure function is written as~\cite{KB:1989,BK:1992, Martin:1999}
\begin{equation}\label{eq:F2}
F_{2}(x,Q^2) = F_2(VMD) + \frac{Q^2}{Q^2 + Q^2_0} F_2^{\mathrm{QCD}}(\bar{x}, Q^2+Q^2_0),
\end{equation}
where 
\begin{equation}
\bar{x} = \frac{Q^2 + Q^2_0}{s+Q^2 + Q^2_0 - M^2} = x \frac{Q^2 + Q^2_0}{Q^2 + x Q^2_0}.
\end{equation}
$Q_0$ should be larger than the mass of the heaviest vector meson included in the VMD contribution, 
but smaller than the mass of the lightest vector meson not included. 
While some analysis gave relatively small values of $Q_0^2$ by treating it as free parameter~\cite{H1:1997, Szczurek:2000}, 
it is reasonable to choose $Q_0^2$ in the range $1.0 \sim 1.5~{\rm GeV}^2$~\cite{Martin:1999}.

Various parametrizations of PDFs in the literature may be used for the QCD component, $F_2^{\rm QCD}$.
One of our aims is to make a comparison between the moments of the PDFs extracted from this two-component model and the lattice simulations.
Here we take the CJ15 formalism~\cite{CJ15} and redetermine some of the parameters by fitting data on the proton deep inelastic scattering structure functions. 
The second moments of the leading twist PDFs are also derived. 
 
\section{Vector Meson Dominance}
The vector meson dominance term has the form
\begin{equation}
F_2(VMD) = \frac{Q^2}{\pi} \sum_V \frac{M^4_V \sigma_{VN}}{f^2_V (Q^2 + M_V^2)^2},
\end{equation}
where $V=\rho^0, \omega$ and $\phi$ and the photon-vector-meson coupling constants are
\begin{equation}
\frac{f_V^2}{4\pi} = \frac{\alpha^2 M_V}{3 \Gamma_{V\rightarrow e^+ e^-}},
\end{equation}
equal to $2.28$, $26.14$, and $14.91$ for $\rho^0$, $\omega$ and $\phi$, respectively.
For the vector meson-proton cross sections, we take
\begin{eqnarray}
\sigma_{\rho p} &=& \sigma_{\omega p} = \frac{1}{2} \Big[ \sigma(\pi^+ p) + \sigma(\pi^- p) \Big],\nonumber\\
\sigma_{\phi p} &=& \sigma( K^+ p) + \sigma(K^- p) - \frac{1}{2} \Big[ \sigma(\pi^+ p) + \sigma(\pi^- p) \Big],
\end{eqnarray}
together with the parametrisation form~\cite{DL:1992}
\begin{eqnarray}\label{eq:sigma-vp}
\sigma_{\rho p} &=& \sigma_{\omega p} = 13.63 s^{\epsilon} + 31.79 s^{-\eta},\nonumber\\
\sigma_{\phi p} &=& 10.01 s^{\epsilon} + 2.72 s^{-\eta}, 
\end{eqnarray}
where $\epsilon=0.08$ and $\eta=0.45$ are taken from Regge theory and the resulting cross sections are in unit of mb.

The VMD contribution is a higher twist (HT) effect because of the vector meson propagators, so that it is negligible for large $Q^2$.
Moreover, it is only relevant when the lifetime of the hadronic fluctuation of photon is larger than the interaction time $\tau_{\rm int} \sim R$~\cite{Levy:1997}, 
\begin{equation}
\tau \sim \frac{1}{\Delta E} \ge R,
\end{equation}
where $R$ is the electromagnetic radius of the proton and in the target reference frame
\begin{equation}
\Delta E = \frac{M_V^2 + Q^2}{Q^2} \cdot M_N x.
\end{equation}
This constraint usually implies that the VMD contributions are important in region where Bjorken $x$ is small, $x \le 0.1$.
In order to incorporate this effect, the standard VMD component was modified by introducing a form factor in~\cite{Szczurek:2000},
\begin{equation}
F_{2}(VMD) = \frac{Q^2}{\pi} \sum_V \frac{M^4_V \sigma_{VN}}{f^2_V (Q^2 + M_V^2)^2} \Omega(x,Q^2),
\end{equation}
where a Gaussian form was preferred by the best fit, 
\begin{equation}
\Omega(x,Q^2) = \exp ( - (\Delta E/ \lambda_G)^2 ),
\end{equation}
with $\lambda_G = 0.50 \mathrm{GeV}$. However, with this choice, the VMD contributions survive up to relatively large $x \sim 0.5$.
In fact, the proton radius provides a natural characteristic scale and therefore in our analysis we choose instead
\begin{equation}
\lambda_G = 1/R = 0.25~{\rm GeV}.
\end{equation}
A possible non-diagonal term corresponding to the $\rho^0 p \rightarrow \omega^0 p$ transition was found to be negligibly small~\cite{Kwiecinski:1983}, 
while the non-diagonal $\omega p \rightarrow \phi p$ and $\rho p \rightarrow \phi p$ transitions are neglected on the basis of the Zweig rule. 

\section{The QCD component}
The standard partonic parametrization, together with $1/Q^2$ corrections such as target mass correction (TMC) and QCD induced higher twist effects~\cite{CJ15}, 
could be applied to the QCD component of the proton structure function,
\begin{equation}\label{eq:F2-QCD}
F^{{\rm QCD}}_2(x,Q^2) = F_2^{\rm LT}(x,Q^2) \left( 1 + \frac{C_{\rm HT}(x)}{Q^2} \right),
\end{equation}
where $F_2^{\rm LT}$ denotes the leading-twist structure function, including TMC effects,
\begin{eqnarray}
F_2^{\rm LT}(x,Q^2) &=& \frac{(1+\rho)^2}{4\rho^3} F_2^{(0)}(\xi,Q^2) + \frac{3x(\rho^2 -1)}{2\rho^4} \int_{\xi}^1 du \nonumber\\
&& \times \Big[ 1 + \frac{\rho^2 -1}{2 x \rho}(u-\xi) \Big] \frac{F_2^{(0)}(u,Q^2)}{u^2},
\end{eqnarray} 
with $F_2^{(0)}$ being the structure function in the $M^2/Q^2 \rightarrow 0$ limit, and the modified scaling variable
\begin{equation}
\xi = \frac{2x}{1+\rho}, \ \ \ \rho^2 = 1 + \frac{4 x^2 M^2}{Q^2}.
\end{equation}
In order to allow flexibility in the shape of the higher-twist contribution, following the discussions 
in~\cite{CJ:10}, the higher-twist coefficient function is parametrized by a polynomial function as
\begin{equation}
C_{\rm HT}(x) = h_0 x^{h_1} (1 + h_2 x).
\end{equation}
The initial PDFs  are taken from the CJ15 parametrisation at $m_c^2 = 1.69~{\rm GeV}^2$,
\begin{equation}
x f(x,Q^2) = a_0 x^{a_1} (1-x)^{a_2} (1 + a_3 \sqrt{x} + a_4 x),
\end{equation}
for the valence $u_v = u - \bar{u}$ and $d_v = d - \bar{d}$,  the light antiquark sea $\bar{u} + \bar{d}$, and the gluon distribution $g$.
In~\cite{CJ15}, the $d_v$ distribution was modified by adding a small admixture of the valence $u$-quark PDF with two additional parameters $b$ and $c$,
\begin{equation}
d_v \rightarrow a_0^{d_v} \left( \frac{d_v}{a_0^{d_v}} + b x^c u_v \right).
\end{equation}
In the $x \rightarrow 1$ limit, it leads to a finite, nonzero value of the ratio
\begin{equation}\label{eq:dv-uv-CJ15}
\frac{d_v}{u_v} \rightarrow a_0^{d_v} b = 8.89 \times 10^{-2}.
\end{equation}
The functional form of $\bar{d}/\bar{u}$ at the input scale is taken to be
\begin{equation}
\frac{\bar{d}}{\bar{u}} = a_0 x^{a_1} (1-x)^{a_2} + 1 + a_3 x (1-x)^{a_4} \, .
\end{equation}
The strange quark distribution is related to the light quark sea through a fixed ratio
\begin{equation}
\kappa = \frac{s + \bar{s}}{\bar{u} + \bar{d}},
\end{equation}
which was set to be $0.4$ in~\cite{CJ15}.
The sensitivity of the fit to this parameter was examined by varying $\kappa$ in the range $0.3 \sim 0.5$.

\section{Results}
\label{sec:results}
In contrast to global fits, the present DIS only analysis is subject to some limitations. 
In particular, the present work should be regarded as exploratory, aimed at investigating whether a full scale search based on this approach would be justified. Bearing in mind the limited data set used here we find it necessary to constrain some parameters.  
First of all, the strongest constraint on $\bar{d}/\bar{u}$ comes from the Drell-Yan process, 
so we fix the parameters of $\bar{d}/\bar{u}$ at CJ15 values.
Second, since the proton structure function is less sensitive to the $d$ quark distribution, we fix the $d/u$ ratio from the CJ15 analysis. 
Then, the valence $d$ quark distribution can be expressed as
\begin{equation}\label{eq:d_v}
d_v(x,Q^2) = \left( \frac{d}{u} \right)_{CJ15} u(x,Q^2) - \left( \frac{\bar{d}}{\bar{u}} \right)_{CJ15} \bar{u}(x,Q^2).
\end{equation}
Moreover, we also fix $a_2$, $a_3$, and $a_4$ of $xg$, as the structure function $F_2$ is insensitive to the large $x$ behaviour of gluon distribution.

The parameters $a_0$ for $x u_v$, $x(\bar{d}+\bar{u})$ and $xg$ are constrained by number and momentum sum rules 
and are therefore not free parameters.
There remain 9 free parameters, which will be determined by fitting the DIS data of the proton structure function. 
The experimental data are taken from BCDMS~\cite{BCDMS:1989}, SLAC~\cite{SLAC:1992}, NMC~\cite{NMC:1997}, E665~\cite{E665:1996}. 
While the CJ15 analysis fits data above $x_{\rm min} = 5 \times 10^{-3}$, 
we also include H1 data in the very small $x$ region~\cite{H1:1996}, down to $1.3 \times 10^{-4}$.
We choose data points of several $Q^2$ bins around $5$, $10$, $20$, $25$ and $50~{\rm GeV}^2$,
which are sufficient to fix the LT PDF parameters as well as the HT coefficient function.
Statistical and systematic errors are added in quadrature.

The LT structure function  in (\ref{eq:F2-QCD}), $F_2^{\rm LT}$, is derived by NLO QCD evolution using APFEL program~\cite{APFEL},
 including target mass corrections. 
By fixing $\kappa = 0.4$, the parameters of the initial PDFs and the HT coefficient function are given in Tables~\ref{tab:Fit-1.0} and \ref{tab:Fit-1.5},
corresponding to $Q_0^2 = 1.0~{\rm GeV}^2$ and $Q_0^2 = 1.5~{\rm GeV}^2$, respectively.
Compared with the CJ15 results, the parameters $a_0$ and $a_1$ for the gluon distribution are much smaller, 
while the HT coefficient function is enhanced in the large $x$ region, because of the larger value of $h_2$.

\begin{table*}
\caption{\label{tab:Fit-1.0} The fitted LT parameter values for $u_v$, $\bar{d} + \bar{u}$, and $g$ PDFs from NLO analysis, 
and the HT parameters for the coefficient function, with $Q^2_0 = 1.0~{\rm GeV}^2$ and $\kappa = 0.4$.
The $\chi^2_{d.o.f} = 116/(129-9) = 0.97$. }
\begin{indented}
\item[]\begin{tabular}{@{}cccccc}
\br
   {\rm LT}    &               $u_v$                   &   $\bar{d} + \bar{u}$              &                   $g$                        & {\rm HT}  &    \      \\ 
\mr
 $a_0$        &              $2.4525$               &             $0.11871$               &             $23.617$                       &   $h_0$   &    $-1.2244 \pm 0.37427$       \\       
 $a_1$         &  $0.61931 \pm 0.0048790$  &  $-0.20943 \pm 0.055740$   &    $0.36755 \pm 0.035769$   &   $h_1$   &    $0.59992 \pm 0.12744$       \\ 
 $a_2$         &  $3.6177 \pm 0.032649$    &             $8.3286 $                &               $6.4812$                    &   $h_2$   &    $-9.5558 \pm 2.7204$        \\  
 $a_3$        &                  $0$                    &                   $0$                      &             $-3.3064$                    &      \         &        \    \\  
 $a_4$         &  $3.5445 \pm 0.13391$     &  $26.215 \pm 7.5988$          &               $3.1721$                    &      \         &        \     \\
\br
\end{tabular}
\end{indented}
\end{table*}
\begin{table*}
\caption{\label{tab:Fit-1.5} The fitted LT parameter values for $u_v$, $\bar{d} + \bar{u}$, and $g$ PDFs from NLO analysis, 
and the HT parameters for the coefficient function, with $Q^2_0 = 1.5~{\rm GeV}^2$ and $\kappa = 0.4$.
The $\chi^2_{d.o.f} = 138/(129-9) = 1.15$. }
\begin{indented}
\item[]\begin{tabular}{@{}cccccc}
\br
   {\rm LT}    &               $u_v$                   &   $\bar{d} + \bar{u}$              &                   $g$                        & {\rm HT}  &    \      \\ 
\mr
 $a_0$        &              $1.7831$               &             $0.29270$               &             $16.532$                       &   $h_0$   &    $-0.87901 \pm 0.37427$       \\       
 $a_1$         &  $0.54552 \pm 0.010214$  &  $-0.049054 \pm 0.058255$   &    $0.25347 \pm 0.030693$   &   $h_1$   &    $0.81821 \pm 0.065805$       \\ 
 $a_2$         &  $3.6998 \pm 0.030355$    &             $8.3286 $                &               $6.4812$                    &   $h_2$   &    $-23.867 \pm 5.9689$        \\  
 $a_3$        &                  $0$                    &                   $0$                      &             $-3.3064$                    &      \         &            \\  
 $a_4$         &  $5.3979 \pm 0.19655$     &  $14.771 \pm 4.6105$          &               $3.1721$                    &      \         &           \\ 
\br
\end{tabular}
\end{indented}
\end{table*}
\begin{figure}[ht]
\begin{center}
\includegraphics[width=\columnwidth]{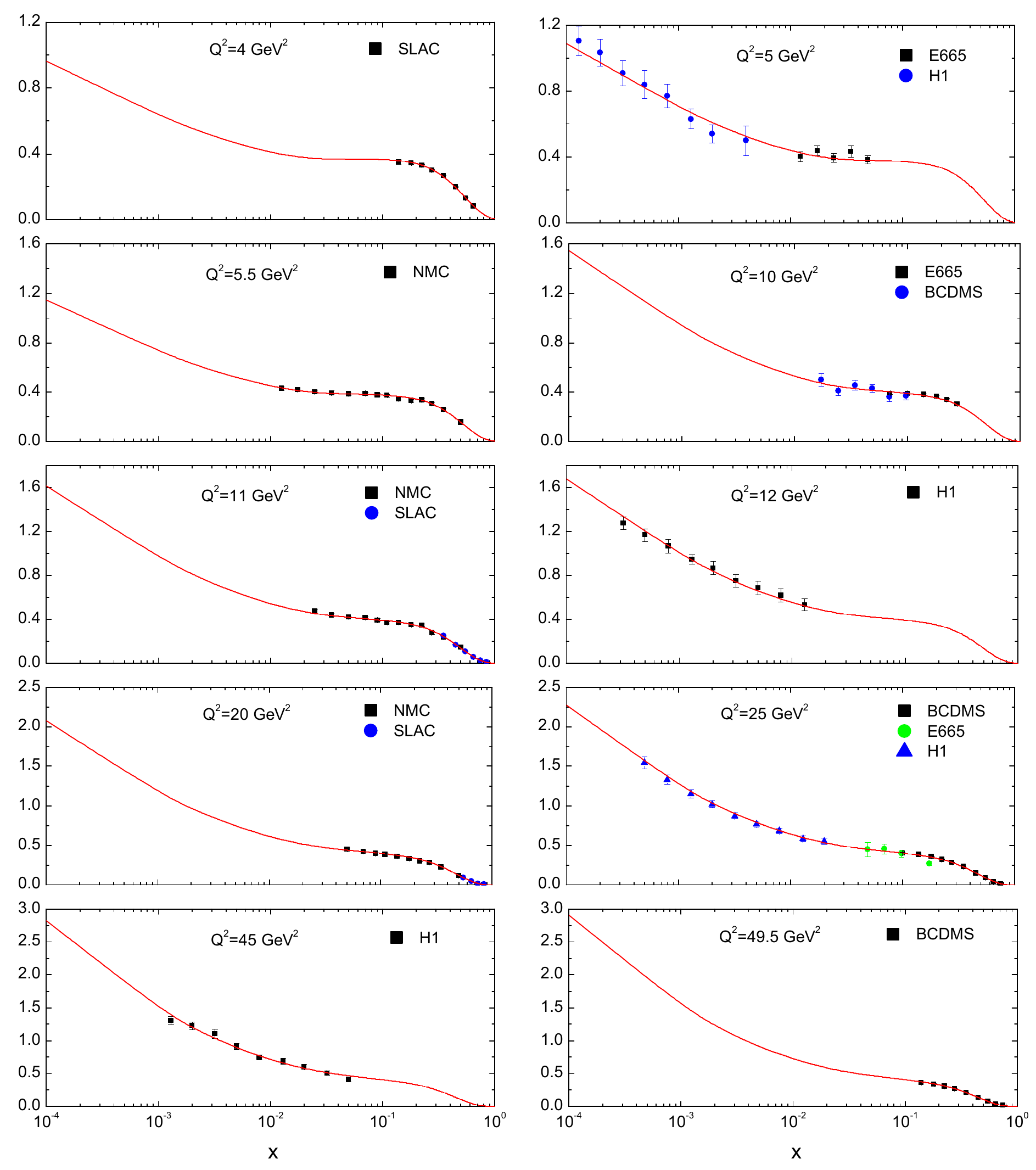}
\caption{The fit results of the proton structure functions with $Q_0^2 = 1.0~{\rm GeV}^2$ and $\kappa = 0.4$. }
\label{fig:F2p}
\end{center}
\end{figure}
\begin{figure}[ht]
\includegraphics[width=\columnwidth]{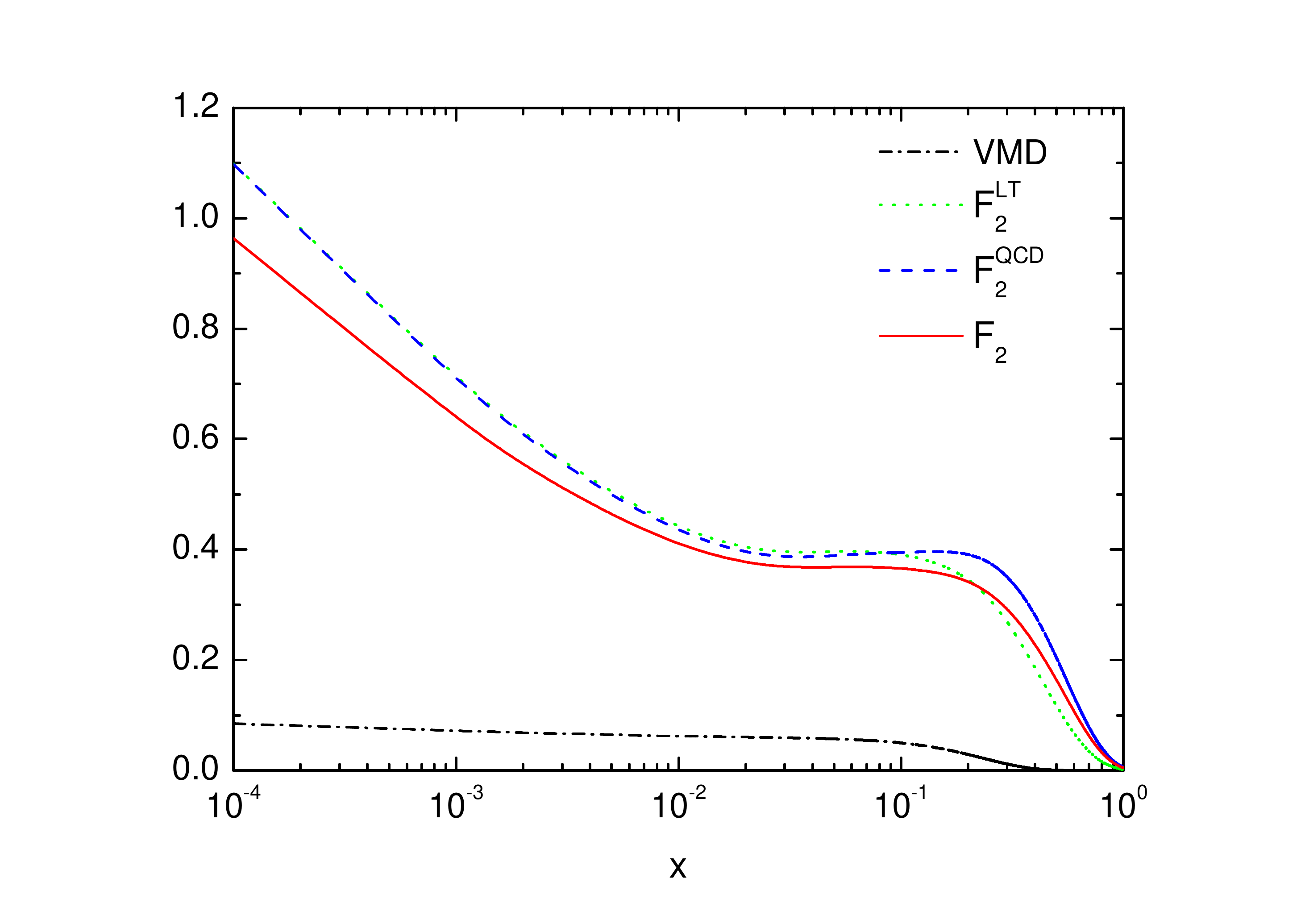}
\caption{Individual contribution from each component to the proton structure function at $Q^2 = 4~{\rm GeV}^2$, with $Q_0^2 = 1.0~{\rm GeV}^2$ and $\kappa = 0.4$. }
\label{fig:F2-4GeV2}
\end{figure}

In the case of $Q_0^2 = 1.0~{\rm GeV}^2$ and $\kappa = 0.4$, the fitted structure functions are shown in Fig.~\ref{fig:F2p},  with $\chi^2_{d.o.f} = 116/(129-9) = 0.97$.
For $Q^2 = 4~{\rm GeV}^2$, we also show in Fig.~\ref{fig:F2-4GeV2} the individual contribution from each component to the total structure function. 
Although negligible for large $Q^2$, the VMD contribution accounts for $(10-20)\%$ of the total $F_2$ at $x \le 0.1$.
The significant enhancement in $F^{\rm QCD}_2$ compared to $F_2^{\rm LT}$ when $x > 0.1$ indicates large HT effect, which is associated with the parameter $h_2$.
\begin{figure}[ht]
\includegraphics[width=\columnwidth]{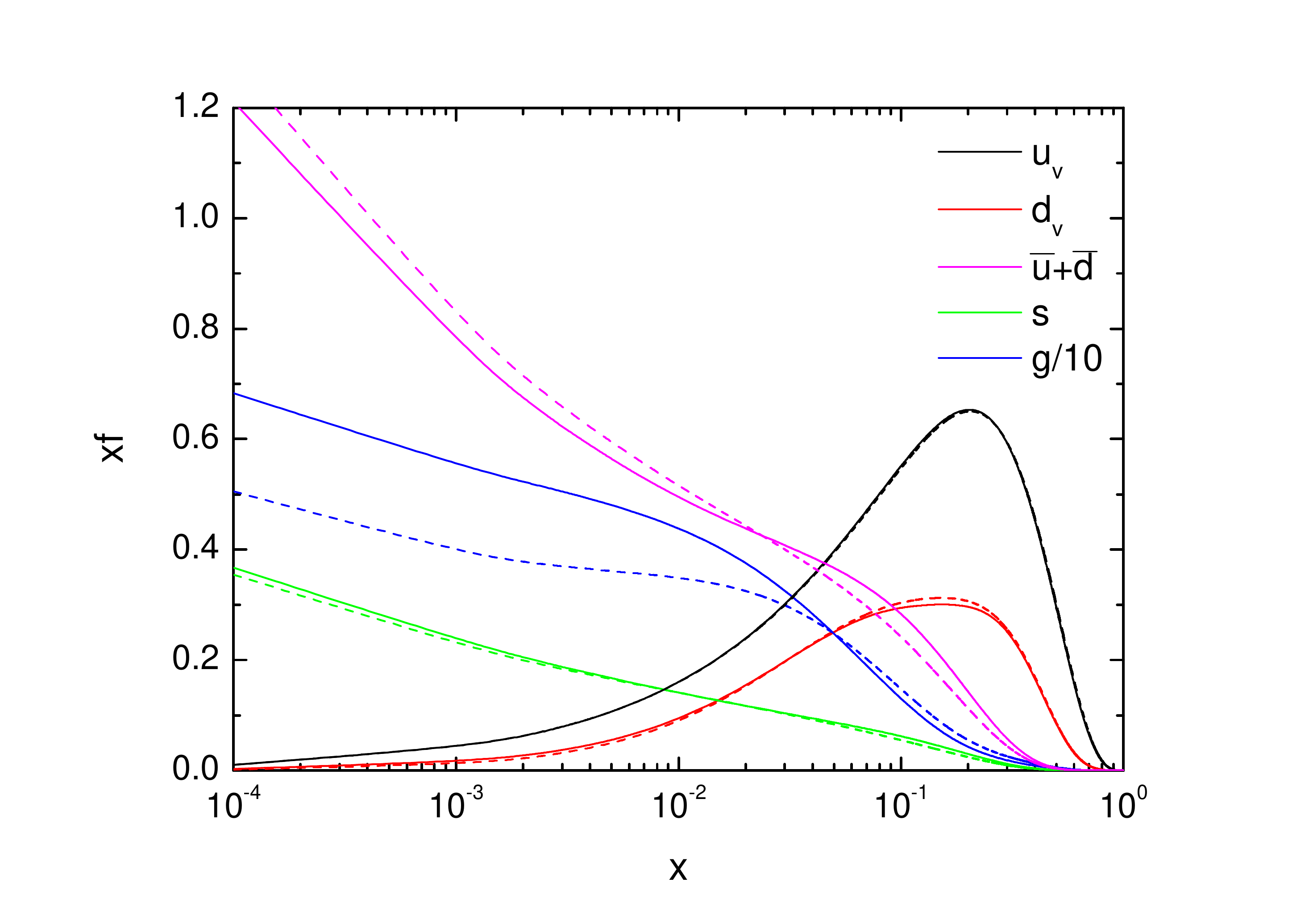}
\caption{Parton distribution functions at $Q^2 = 4~{\rm GeV}^2$. The solid lines are our fit results, while the dashed lines are the corresponding PDFs from CJ15 results.}
\label{fig:xf-4GeV2}
\end{figure}

The  LT PDFs at $Q^2 = 4~{\rm GeV}^2$ are displayed in Fig.~\ref{fig:xf-4GeV2} and compared with the CJ15 results.
The valence quark distributions $x u_v$ and $x d_v$ are almost unchanged. 
In addition, the $d_v$ given in~(\ref{eq:d_v}) results in $d_v/u_v \rightarrow 8.80 \times 10^{-2}$ as $x \rightarrow 1$, 
which is consistent with the CJ15 value in~(\ref{eq:dv-uv-CJ15}).
However, there is a noticeable difference for $x(\bar{u}+\bar{d})$ and a large difference for the distribution $xg$.
Our analysis tends to result in a smaller sea quark distribution and a significantly larger gluon distribution in the small $x$ region.
While in the range $0.05 \le x \le 0.5$, we find a sizeable increase in $x(\bar{u}+\bar{d})$ and small decrease in $xg$, in comparison with the CJ15 PDFs.

We also repeat the same fit procedure with different values of $Q^2_0$ and $\kappa$. The corresponding second moments of the LT PDFs at $Q^2 = 4~{\rm GeV}^2$ are summarized in Table~\ref{tab:moments}.
Within the limits of $Q_0^2$ and $\kappa$, the momentum fraction carried by gluons is a few percent smaller than the previous global-fit determination.
\begin{table*}
\caption{\label{tab:moments} The second moments of LT PDFs at $Q^2 = 4~{\rm GeV}^2$. }
\begin{indented}
\item[]\begin{tabular}{@{}cccccc}
\br
   $Q_0^2 ({\rm GeV}^2)$  &        $\kappa$    &    $\langle x \rangle_{u^+}$    &   $\langle x \rangle_{d^+}$    &     $\langle x \rangle_{s^+}$       &       $\langle x \rangle_g$   \\ 
\mr
                       \                  &              $0.3$    &             $0.3552$                    &           $0.2049$                    &                $0.0283$                     &                $0.4033$       \\       
                   $1.0$              &              $0.4$    &             $0.3543$                    &           $0.2039$                    &                $0.0344$                     &                $0.3993$             \\ 
                       \                  &              $0.5$    &             $0.3538$                    &           $0.2027$                    &                $0.0397$                     &                $0.3958$        \\  
\mr
                       \                  &              $0.3$    &             $0.3572$                    &           $0.2108$                    &                $0.0312$                     &                $0.3920$          \\  
                   $1.5$              &              $0.4$    &             $0.3554$                    &           $0.2095$                    &                $0.0386$                     &                $0.3879$           \\ 
                       \                  &              $0.5$    &             $0.3542$                    &           $0.2084$                    &                $0.0459$                     &                $0.3831$           \\ 
\mr
   {\rm CJ15}~\cite{CJ15}  &              $0.4$    &             $0.3488$                    &           $0.1965$                    &                $0.0311$                     &                $0.4152$        \\      
\br
\end{tabular}
\end{indented}
\end{table*}
%


\section{Conclusion}
In this paper, we argue that a two-component model of nucleon structure functions incorporating the VMD contribution as well as the  
correct photo-production limit is more appropriate when describing DIS data in the low $Q^2$ and small $x$ region.
The QCD component is expressed in terms of the CJ15 parametrization.
The parameters of the LT PDFs and the HT coefficient function were redetermined by fitting  inclusive deep-inelastic scattering data for the proton. 
Our analysis shows that the application of the two component model mainly affects the $x (\bar{u} + \bar{d})$ and $xg$ distributions in the small and moderate $x$ regions.
The second moments of the PDFs were also derived, with the results 
suggesting only small increases in the moments of the quark distributions and correspondingly a smaller value of the momentum fraction carried by gluons.
Far from explaining the significant discrepancy between the global-fit determinations and the lattice results, the present analysis confirms the former. 

Further improvements can be made by embedding this two-component model of structure functions into a global-fit program,  
in which one can remove the constraints used in the present work and determine all of the parameters more precisely.

\ack{
We would like to thank Wally Melnitchouk for helpful discussions and comments . 
This work was supported by the Australian Research Council through  Discovery Projects DP151103101 and DP180100497.
}


\section*{References}

\begin{thebibliography}{55}
%
\bibitem{MRST:1998}
Martin A D, Roberts R G, Stirling W G and Thorne R S 1998 Parton distributions: a new global analysis {\it Eur. Phys. J.} C {\bf 4} 463

\bibitem{GRV:1998}
Gl\"{u}ck M, Reya E and Vogt A 1998 Dynamical parton distributions revisited {\it Eur. Phys. J.} C {\bf 5} 461

\bibitem{MMHT14}
Harland-Lang L A, Martin A D, Motylinski P and Thorne R S 2015 Parton distributions in the LHC era: MMHT 2014 PDFs {\it Eur. Phys. J.} C {\bf 75} 204

\bibitem{NNPDF:2015}
Ball R D {\it et al} (NNPDF Collaboration) 2015 Parton distributions for the LHC Run II {\it J. High Eenrgy Phys.} JHEP04(2015)040

\bibitem{CJ15}
Accardi A, Brady L T, Melnitchouk W, Owens J F and Sato N 2016 Constraints on large-$x$ parton distributions from new weak boson production and deep-inelastic scattering data 
{\it Phys. Rev.} D {\bf 93} 114017

\bibitem{Ji:2013}
Ji X 2013 Parton physics on a Euclidean lattice {\it Phys. Rev. Lett.} {\bf 110} 262002

\bibitem{Lin:2015}
Lin H W, Chen J W, Cohen S D and Ji X 2015 Flavor structure of the nucleon sea from lattice QCD {\it Phys. Rev.} D {\bf 91} 054510

\bibitem{Radyushkin:2017}
Radyushkin A V 2017 Quasi-parton distribution functions, momentum distributions, and pseudo-parton distribution functions {\it Phys. Rev.} D {\bf 96} 034025

\bibitem{Orginos:2017}
Orginos K, Radyushkin A, Karpie J and Zafeiropoulos S 2017 Lattice QCD exploration of parton pseudo-distribution functions {\it Phys. Rev.} D {\bf 96} 094503

\bibitem{Lin:2018}
Lin H W {\it et al} 2018 Parton distributions and lattice QCD calculations: A community white paper {\it Prog. Part. Nucl. Phys.} {\bf 100} 107

\bibitem{Alexandrou:2017}
Alexandrou C {\it et al} 2017 Nucleon spin and momentum decomposition using lattice QCD simulations {\it Phys. Rev. Lett.} {\bf 119} 142002

\bibitem{Braun:1995}
Braun V, Gornicki P and Mankiewicz L 1995 Ioffe-time distributions instead of parton momentum distributions in description of deep inelastic scattering 
{\it Phys. Rev.} D {\bf 51} 6036

\bibitem{Vanttinen:1998}
V\"{a}nttinen M, Piller G, Mankiewicz L, Weise W and Eskola K J 1998 Nuclear quark and gluon distributions in coordinate space {\it Eur. Phys. J.} A {\bf 3} 351 

\bibitem{KB:1989}
Kwiecinski J and Badelek B 1989 Analysis of the electroproduction structure functions in the low $Q^2$ region combining the vector meson dominance
and the parton model with possible scaling violation {\it Z. Phys.} C {\bf 43} 251

\bibitem{BK:1992}
Badelek B and Kwiecinski J 1992 Electroproduction structure function $F_2$ in the low $Q^2$, low $x$ region {\it Phys. Lett.} B {\bf 295} 263

\bibitem{Martin:1999}
Martin A D, Ryskin M G , Stasto A M 1999 The description of $F_2$ at low $Q^2$ {\it Eur. Phys. J.} C {\bf 7} 643

\bibitem{H1:1997}
Adloff C {\it et al} (H1 Collaboration) 1997 A measurement of the proton structure function $F_2(x,Q^2)$ at low $x$ and low $Q^2$ at HERA {\it Nucl. Phys.} B {\bf 497} 3

\bibitem{Szczurek:2000}
Szczurek A and Uleshchenko V 2000 Nonpartonic components in the nucleon structure functions at small $Q^2$ in a broad range of $x$ {\it Eur. Phys. J.} C {\bf 12} 663 

\bibitem{DL:1992}
Donnachie A and Landshoff P V 1992 Total cross sections {\it Phys. Lett.} B {\bf 296} 227

\bibitem{Levy:1997}
Levy A 1997 The proton and photon, who is probing whom? {\it Phys. Lett.} B {\bf 404} 369

\bibitem{Kwiecinski:1983}
Kwiecinski J 1983 Linking total cross-sections for hard and soft processes through analyticity in $Q^2$ and sum rules {\it Phys. Lett.} B {\bf 120} 418

\bibitem{CJ:10}
Accardi A, Christy M E, Keppel C E, Melnitchouk M, Monaghan P, Morfin J G, and Owens J F 2010 New parton distributions from large-$x$ and low-$Q^2$ data {\it Phys. Rev.} D {\bf 81} 034016
\bibitem{BCDMS:1989}
Benvenuti A C {\it et al} (BCDMS Collaboration) 1989 A high statistics measurement of the proton structure functions $F_2(x,Q^2)$ and R from deep inelastic muon scattering at high $Q^2$ {\it Phys. Lett.} B {\bf 223} 485

\bibitem{SLAC:1992}
Whitlow L W, Riordan E M, Dasu S, Rock S and Bodek A 1992 Precise measurements of the proton and deuteron structure functions 
from a global analysis of the SLAC deep electron scattering cross-sections {\it Phys. Lett.} B {\bf 282} 475

\bibitem{NMC:1997}
Arneodo M {\it et al} (New Muon Collaboration) 1996 Measurement of the proton and deuteron structure functions, $F_2^p$ and $F_2^d$, and of the ratio $\sigma_L/\sigma_T$ {\it Nucl. Phys.} B {\bf 483} 3

\bibitem{E665:1996}
Adams M R {\it et al} (E665 Collaboration) 1996 Proton and deuteron structure functions in muon scattering at $470~{\rm GeV}$ {\it Phys. Rev.} D {\bf 54} 3006

\bibitem{H1:1996}
Aid S {\it et al} (H1 Collaboration) 1996 A measurement and QCD analysis of the proton structure function $F_2(x,Q^2)$ at HERA {\it Nucl. Phys.} B {\bf 470} 3

\bibitem{APFEL}
Bertone V, Carrazza S and Rojo J 2013 APFEL: A PDF evolution library with QED corrections {\it Comput. Phys. Commun.} {\bf 185} 1647
%
\end{thebibliography}
\end{document}